\documentclass[aps,pre,twocolumn,superscriptaddress,showpacs,longbibliography]{revtex4-1}

\newcommand{\gapprox}{%
\mathrel{%
\setbox0=\hbox{$>$}
\raise0.6ex\copy0\kern-\wd0
\lower0.65ex\hbox{$\sim$}
}}

\usepackage{graphicx}
\usepackage{grffile}

\usepackage{soul}
\graphicspath{{im/}}

\usepackage{amsmath}
\usepackage{latexsym}
\usepackage{comment}
\usepackage{bm}
\usepackage{gensymb}
\usepackage{hyperref}
\usepackage[british]{babel}
\usepackage{url}
\usepackage{color}
\usepackage[normalem]{ulem}
\usepackage{mathtools}
\hypersetup{colorlinks,urlcolor=blue}

\begin{document}
\title{Homogeneous and heterogeneous populations of active rods in two-dimensional channels}
\date{\today}
\author{V. Khodygo} 
\email{vlk@aber.ac.uk}
\affiliation{Institute of Biological, Environmental and Rural Sciences, Aberystwyth University, Penglais Campus, Aberystwyth, Ceredigion, Wales, SY23 3DA, UK}
\affiliation{Institute of Mathematics, Physics and Computer Science, Aberystwyth University, Penglais Campus, Aberystwyth, Ceredigion, Wales, SY23 3DB, UK}
\author{M. T. Swain}
\thanks{Swain and Mughal, as senior authors, contributed equally to this work.}
\affiliation{Institute of Biological, Environmental and Rural Sciences, Aberystwyth University, Penglais Campus, Aberystwyth, Ceredigion, Wales, SY23 3DA, UK}
\author{A. Mughal}
\thanks{Swain and Mughal, as senior authors, contributed equally to this work.}
\affiliation{Institute of Mathematics, Physics and Computer Science, Aberystwyth University, Penglais Campus, Aberystwyth, Ceredigion, Wales, SY23 3DB, UK}

\begin{abstract}
Active swarms, consisting of individual agents which consume energy to move or produce work, are known to generate a diverse range of collective behaviors. 
Many examples of active swarms are biological in nature (e.g., fish shoals and bird flocks) and have been modeled extensively by numerical simulations. 
Such simulations of swarms usually assume that the swarm is homogeneous; that is, every agent has exactly the same dynamical properties. 
However, many biological swarms are highly heterogeneous, such as multispecies communities of micro-organisms in soil, 
and individual species may have a wide range of different physical properties. 
Here we explore heterogeneity by developing a simple model for the dynamics of a swarm of motile heterogeneous rodlike bacteria in the absence of hydrodynamic effects. 
Using molecular dynamics simulations of active rods confined within a two-dimensional rectangular channel, 
we first explore the case of homogeneous swarms and show that the key parameter governing both dynamics is ratio of the motility force to the steric force. 
Next we explore heterogeneous or mixed swarms in which the constituent self-propelled rods have a range of motilities and steric interactions. 
Our results show that the confining boundaries play a strong role in driving the segregation of mixed populations. 
\end{abstract}
\maketitle

\section{Introduction}\label{sec:intro}

Active swarms are composed of individual self-propelled agents that are capable of converting energy into motion. 
Natural examples include bacterial swarms \cite{Lushi2014}, bird flocks \cite{Ballerini2008},  fish shoals \cite{Hager1991}, and mammalian herds \cite{King2012}. 
Anthropogenic examples include the behavior of crowds \cite{Silverberg2013,Karamouzas2014} and the flow of traffic \cite{Helbing2001}. 
In such systems the collective motion of a large number of simple individuals can result in complex nonequilibrium behavior. 
While such phenomena are responsible for some of the most beautiful displays in the natural world,
such as the murmuration of starlings or the collective motion of fish shoals; 
it can also be the source of deep inconvenience, such as, for example, the frustration of being stuck in stop-go traffic.

Recently, there has been considerable interest in the swarming of simple rodlike bacteria (for example, \emph{Bacillus subtilis}) confined to the surface of a two-dimensional interface \cite{Wioland2016, Wioland2013, Kaiser2014b, Cisneros2010}.
Such swarms, captured within a free-standing film or between a solid-solid or solid-liquid interface,
have the advantage that they can be imaged easily by real-space microscopy and provide an ideal environment for the study of active matter.  

In this paper we conduct numerical simulations of self-propelled rods on a (quasi-)two-dimensional [(quasi-)2D]surface.
Such simulations provide an analogy for dense swarms of motile rodlike bacterial swarms on a flat interface.
To prevent rods from overlapping we impose a simple steric interaction between rods:
Each rod is divided into segments and segments from neighboring rods repel each other using a Hookean potential.
The model we are using was first introduced by Peruani \textit{et al.} \cite{Peruani2006} to consider the effects of cell shape. 
It ignores hydrodynamic interactions due to the assumption that the cells are densely packed and move in a very viscous media.
Myxobacteria gliding on a surface is an example of such a system \cite{Peruani2012}, where the lengths of the rodlike bacteria are distributed heterogeneously.
Similar approaches may use a repulsive Yukawa force that has been restricted to only act on overlapping rod segments \cite{Wensink2008}.
The use of a Hookean potential has two advantages --- first, the overlap energy between neighboring rods scales in a simple manner,
and second, by varying the magnitude of the spring constant the interaction between rods can be tuned continuously from soft to hard. 
Self motility is imposed by introducing a constant propulsion force along the axis of the rod. 
Collisions between rods are resolved into a force acting on the center of mass and a torque that acts to change the orientation of the rod. 
The position and orientation of the rods, along with the center of mass forces and torques, are used to perform an overdamped molecular dynamics simulation. 

Despite the simplicity of the model it is capable of reproducing a wide range of behaviors commonly seen in dense bacterial swarms. 
Simulations of this type have provided insights into phenomena such as turbulence in active systems, corporative swarming, and alignment on long length scales \cite{Wensink2012a}. 
In addition, features, such as \textit{hedgehog}like formations (whereby rods near a wall jam together forming a fan-shaped cluster) and giant density fluctuations were also successfully reproduced \textit{in silico} \cite{Fily2014a,Wensink2008} and have been observed in real bacterial communities \cite{Mazza2016, Peruani2012} (and other active systems, e.g., Ref. \cite{Kudrolli2008}.
The method has also been used to simulate directed bacterial transport of mesoscopic carriers \cite{Kaiser2014b}.

We study a swarm of self-propelled rods in a (quasi-)2D channel between two narrowly separated parallel walls. 
A similar confining geometry has been studied experimentally (and by means of simulations) by Wioland \textit{et al.} \cite{Wioland2016}. 
We note that a key difference between the work of Wioland \textit{et al.} and the present study is the effect of hydrodynamic forces, which we do not consider. 
Nevertheless, it was found by Wioland \textit{et al.} that the nature of the flow could vary from turbulent to laminar depending on the width of the channel. 
We show that the presence of the channel boundaries leads to a layering effect whereby rods are densely packed along the channel boundaries, 
with subsequent internal layers forming behind the surface layer. 

Previous studies (e.g., Refs. \cite{Wioland2016, Weitz2015, Wensink2008}) have focused on the highly idealised case of a homogeneous active population. 
These, however, are in contrast with the heterogeneity of natural environments, where multispecies swarms have been likened to moving ecosystems \cite{Ben-Jacob2016}. 
Even within a single species swarm, the cell aspect ratio may change, and this can be important for the ability of bacteria to swarm efficiently \cite{Ilkanaiv2017}. 
To date, typical studies of active matter have involved fully homogeneous populations or at most binary mixtures (e.g., with different chiralities \cite{Chen2015, Mijalkov2013, Ai2015}, motilities \cite{Costanzo2014a, Mones2015}, or shapes \cite{Yang2014a, Allahyarov2018}). 
A notable exception is the study by active polydisperse disks which are found to form glassy states \cite{Fily2014a}. 

Here we directly consider the effect of heterogeneity in a population of active rods.  
We focus on the case where both the driving force for individual rods, and the nature of the steric interaction between any pair of rods, are chosen from a random distribution.
The main finding is that the channel walls drive the segregation of a heterogeneous population,
so that hard rods and strongly driven rods are found in greater concentrations at the channel boundaries.
This finding may have consequences for developing passive systems that can sort or segregate bacterial populations by geometry alone.

This paper is organised as follows. 
In Sec. \ref{sec:theory} we describe our model and numerical scheme. 
Section \ref{sec:hom_systems} contains an analysis of the behavior of homogeneous systems with various packing fractions and steric interactions.
Following this, Secs. \ref{sec:hetero} and \ref{sec:full_hetero} are dedicated to heterogeneous active matter,
where we define a heterogeneous system and investigate its properties by varying the self-propellant forces and steric interactions of the population. 

\section{Theory}\label{sec:theory}

\subsection{Model}

We consider a (quasi-) 2D system in which $N$ active rods are confined within a rectangular channel of length $L$, width $W,$ and area $A=LW$, where $L>W$. 
We define the origin of our coordinate system to lie in the middle of the strip and impose periodic boundary conditions on either side of the strip.
In addition, distance is measured in units of diameter $d$, speed in units of $\mathcal{F}$, time in units of $d/\mathcal{F}$.

For computational ease, the rods are represented as a series of $2M+1$ segments --- i.e., disks of diameter $d$ --- stacked along the long axis of the rod. 
The distance between neighboring segments is $l_o=d/2$ and in all our simulations we fix $M=1$.
Thus all our rods are composed of three segments --- although clearly more segments can be included by increasing the value of $M$; see, for example, Refs. \cite{Wensink2012a, Weitz2015}.

\begin{figure}[h]
\begin{center}
\centering
\includegraphics[width=1.0\columnwidth]{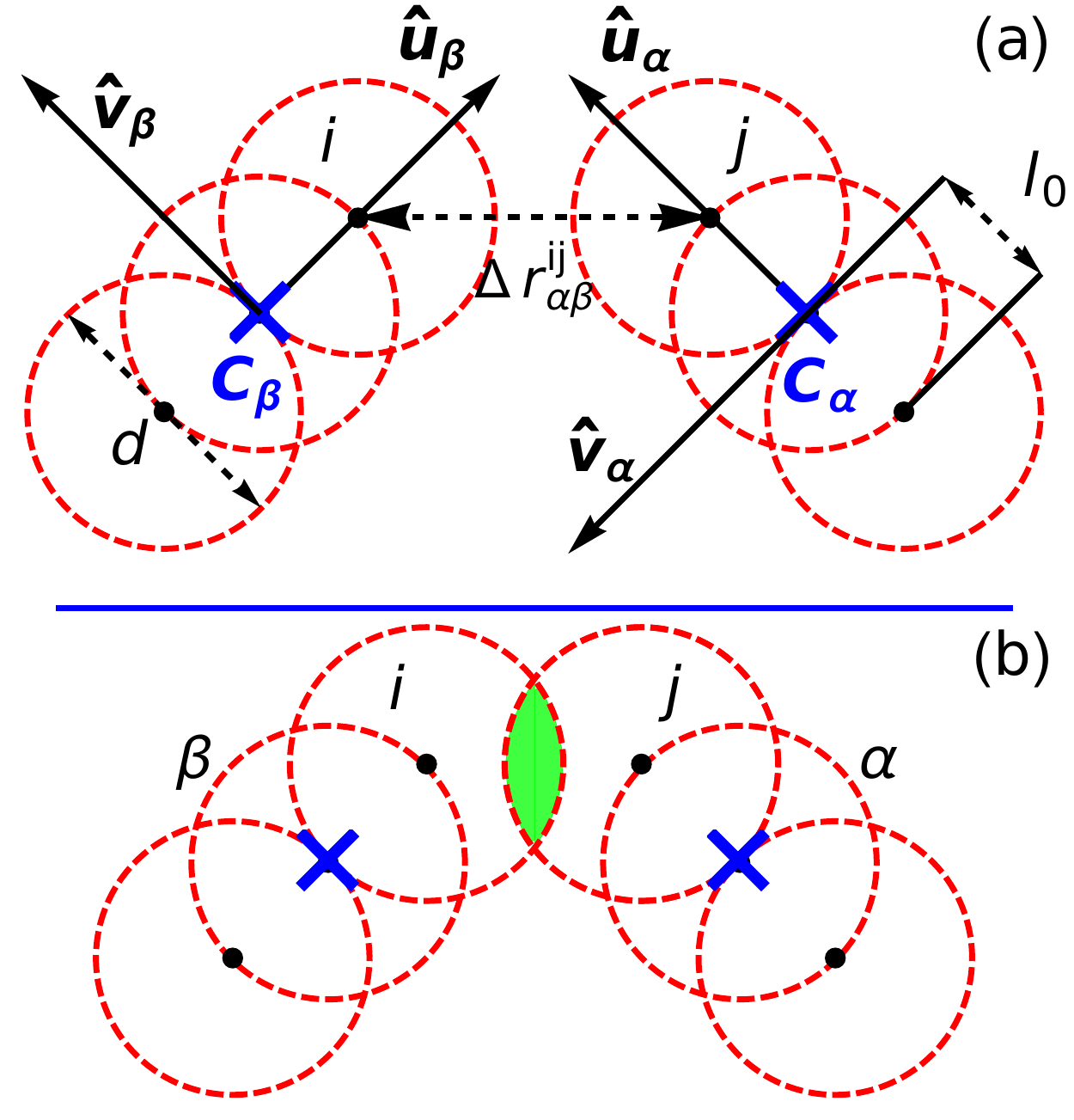}
\caption{
(a) The diagram shows two rods $\alpha$ and $\beta$. 
Each rod is defined by a center of mass (crosses showing ${\bf C_{\alpha}}$ and ${\bf C_{\beta}}$, respectively) and a unit vector (${\bf{\hat{u}}_\alpha}$ and ${\bf{\hat{u}}_\beta}$, respectively) pointing along the forward direction of the rods. 
Each rod is composed of a series of $2M+1$ segments (of diameter $d$) stacked along the unit vector. 
The distance between successive segments is fixed to be $l_0=d/2$. 
A self-propellent force $\mathcal{F}_{\alpha}; \mathcal{F}_{\beta}$ is directed along the corresponding unit vector $\bf{\hat{u}}_\alpha; \bf{\hat{u}}_\beta$ and drives the rod forward. 
(b) If there is an overlap between segments from two different rods (as shown by the shaded (green) region) then there exists a repulsive force between segments where $\Delta r_{\alpha\beta, ij} < d$ between segments $i$ and $j$, as given by  Eq. \ref{eq:spotential}.
}
\label{fig:segment}
\end{center}
\end{figure}

A given rod is specified by its center of mass point ${\bf C}_{\alpha}=(x_{\alpha}(t), y_{\alpha}(t))$ and by its orientation,
given by the unit vector ${\bf{\hat{u}}}_\alpha=(\cos\theta_{\alpha}, \sin\theta_{\alpha})$ [see Fig. \ref{fig:segment}(a)].
In addition, we also define the unit vector ${\bf{\hat{v}}}_\alpha=(-\sin\theta_{\alpha}, \cos\theta_{\alpha})$ perpendicular to ${\bf{\hat{u}}_\alpha}$.
Hence, the coordinates of the $i\text{th}$ segment from the $\alpha\text{th}$ rod is given by, 
\[
{\bf r}_{\alpha, i}={\bf C}_{\alpha}+i l_o{\bf{\hat{u}}_\alpha},
\]
where $i$ is an integer ranging from $[-M, M]$ (so that if $M=1$ we have $i=-1,0,1$). 

Steric interactions between rods are implemented by pairwise interactions between {\it overlapping} segments from different rods; see Fig. \ref{fig:segment}(b). 
We define a repulsive interaction between segment $i$ from rod $\alpha$ with segment $j$ from rod $\beta$ as being given by, 
\begin{equation}
U_{\alpha \beta}^{ij}
 =
\left\{ 
\begin{array}{l l} 
\frac{1}{2}\kappa (d-\Delta r_{\alpha\beta, ij})^2 & \quad \mbox{if $\Delta r_{\alpha\beta,ij}\leq d$}\\
0 & \quad \mbox{if $\Delta r_{\alpha\beta, ij} >d$},\\
\end{array}
\right.
\label{eq:spotential}
\end{equation}
where $\Delta r_{\alpha\beta, ij}=\left| {\bf r}_{\alpha, i}-{\bf r}_{\beta, j} \right|$ is the distance between the centers of the segments,
$\kappa$ is the strength of the interaction and $d$ is the segment diameter. 
Note the interaction energy falls to zero when there is no overlap between the segments thus giving a short-range force.

\begin{figure}[h]
\begin{center}
\centering
\includegraphics[width=1.0\columnwidth]{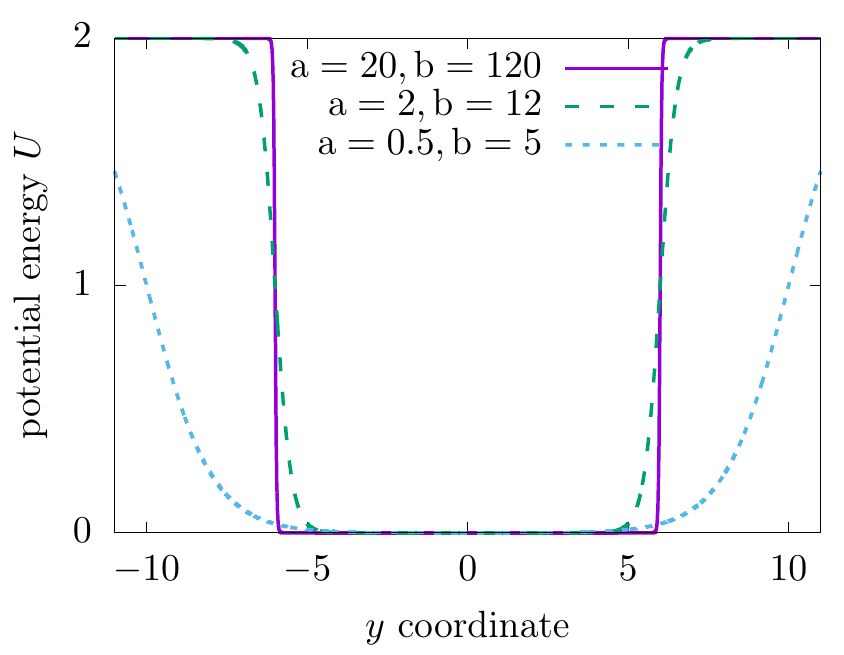}
\caption{
A plot of the the confining potential $U_{\alpha, i}^B(a,b)$ for selected values of $a$ and $b$. 
We plot only the cross-sections since scaling the length $L$ does not bring new effects. 
In our simulations we set $a=20.0$ and $b=120.0$.
}
\label{boundarypotential}
\end{center}
\end{figure}

The rods are prevented from escaping through the walls of the channel (located at $y=\pm W/2$) by a repulsive force as shown in Fig. \ref{boundarypotential}.
This is modelled as a steric interaction between each segment in a given rod and the bounding walls of the channel.
The bounding potential experienced by the $i\text{th}$ segment in the $\alpha\text{th}$ rod is,
\begin{equation}
U_{\alpha, i}^B (a,b)= \tanh(-a y_{\alpha, i} - b) + \tanh(a y_{\alpha, i} - b) +2 
\label{eq:boundary}
\end{equation}
where the width of a channel is defined as $W = 2 b / a$ and both constants $a$ and $b$ define the slope as well. 

Thus, the total force acting on the rod is the sum of the steric forces and the self-propellant forces, and is given by,
\begin{equation}
{\bf F}_{\alpha}
=
-
\sum_{i=-M}^{i=M}
\nabla
U_{\alpha, i} + \mathcal{F}_{\alpha}{\bf{\hat{u}}_\alpha}.
\label{eq:tforce}
\end{equation}
Where the first term in Eq. (\ref{eq:tforce}) accounts for the potential acting on the $i\text{th}$ segment of the $\alpha\text{th}$ rod  due to all the steric interactions, i.e.,
\begin{equation}
U_{\alpha, i}
=
U_{\alpha, i}^B (a,b)
+
\sum_{\alpha \neq \beta}^{N}
\sum_{j=-M}^{j=M}
U_{\alpha \beta}^{ij},
\end{equation}
The second term in Eq. (\ref{eq:tforce}) is the self-propellant force of magnitude $F_{\alpha}$ acting along the direction of the rod, which drives it forward. 
The total force ${\bf F}_{\alpha}$ can be decomposed into a component acting on the center of mass of the rod along the main axis and its orthogonal counterpart:
\begin{equation}
	{\bf F}_{\alpha}^{\text{c.o.m.}}=({\bf F}_{\alpha}\cdot {\bf{\hat{u}}_\alpha}) {\bf{\hat{u}}_\alpha}+({\bf F}_{\alpha}\cdot {\bf{\hat{v}}_\alpha}) {\bf{\hat{v}}_\alpha}.
\end{equation}
In addition, due to asymmetric shape of active rods, forces acting on segments induce a torque:
\begin{equation}
\pmb{\tau}_{\alpha}=
\sum_{i=-M}^{i=M}
\left(
{\bf r}_{\alpha, i}-{\bf C}_{\alpha}
\right) 
\times {\bf F}_{\alpha, i}. 
\end{equation}

\subsection{Numerical Scheme}
We model the dynamics of the rods using overdamped Langevin equations for translational and rotational motion.
Our systems have zero noise --- thus, the dynamics is purely deterministic.
Moreover, since we are interested in collision-induced dynamics only, we omit all hydrodynamic interactions so that active matter is ``dry''. 
Given that, we have:
\begin{eqnarray}
&\eta_{\alpha}^{tr}\dfrac{d{\bf C}_{\alpha}}{dt} &= {\bf F}_{\alpha}^{\text{c.o.m.}},\\
&\eta_{\alpha}^{rot}\dfrac{d\bf{\hat{u}}_{\alpha}}{dt}&=\pmb{\tau}_{\alpha}\times\bf{\hat{u}}_{\alpha},
\label{eq:langevin}
\end{eqnarray} 
where $\eta_{\alpha}^{tr}$ and $\eta_{\alpha}^{rot}$ are the translational and rotational damping constants (tensors) --- see Ref. \cite{Wensink2012a} for details.

We obtain the rod dynamics by using a first-order Euler integration scheme. 
The initial conditions for all simulations is that the rods are placed and orientated randomly and uniformly (i.e., equally distributed) in the channel. 
We run each system for a period of $t_0=10^5$ iterations ($200$ time units, i.e. the transient phase of the simulation)
so that initial effects have died away before continuing the simulation for a longer time period of $5\times 10^6$ over which we collect data and make measurements.

By approximating each rod as a 2D spherocylinder, the packing fraction (i.e., the ratio of the total area of all rods relative to the channel area) is given as \cite{Wensink2012a}, 
\begin{equation}
\label{eq:fraction}
\phi 
= 
\frac{N}{A}
\left( 
2 d M l_0 + \frac{\pi d^2}{4}
\right),
\end{equation}
when there is no overlap between segments from adjacent rods assumed.

In the following simulations we use a channel of length $L = 120 d$ and width $W = 12 d$. 
The resulting area $A$ of a domain is thus given by $L*(W+d)$, where $d$ is a constant term introduced due to the pointlike nature of active rods.
This choice provides us a channel which is sufficiently long for the rods to self-organize into a collective flow and sufficiently narrow as to prevent large scale vorticity from developing. 
In all simulations we consider dense ($\phi=1.1$ to have reliable statistics in every region of the channel) populations of active rods in periodic channels. 

\section{Homogeneous population of rods in a periodic rectangular channel}
\label{sec:hom_systems}
We consider a homogeneous population of rods.
Every rod $\alpha$ has same driving force $\mathcal{F}_{\alpha}=\mathcal{F}$ (where we set $\mathcal{F}=1$), 
each rod segment has the same diameter $d=1$ and hardness $\kappa_{\alpha}=\kappa$.
Hence, the properties of the system depend \textit{almost} solely on the ratio,
\begin{equation}
 \gamma=\dfrac{ d\langle\kappa_{\alpha}\rangle}{\langle\mathcal{F}_{\alpha}\rangle},
 \label{eq:gamma}
\end{equation}
(the exception to this is the confining potential used to simulate the wall boundaries, which does not depend on $\kappa$).
Here we investigate the effect of $\gamma$ on the distribution of rods in the channel.

\begin{figure}[h]
\begin{center}
\centering
\includegraphics[width=1.0\columnwidth]{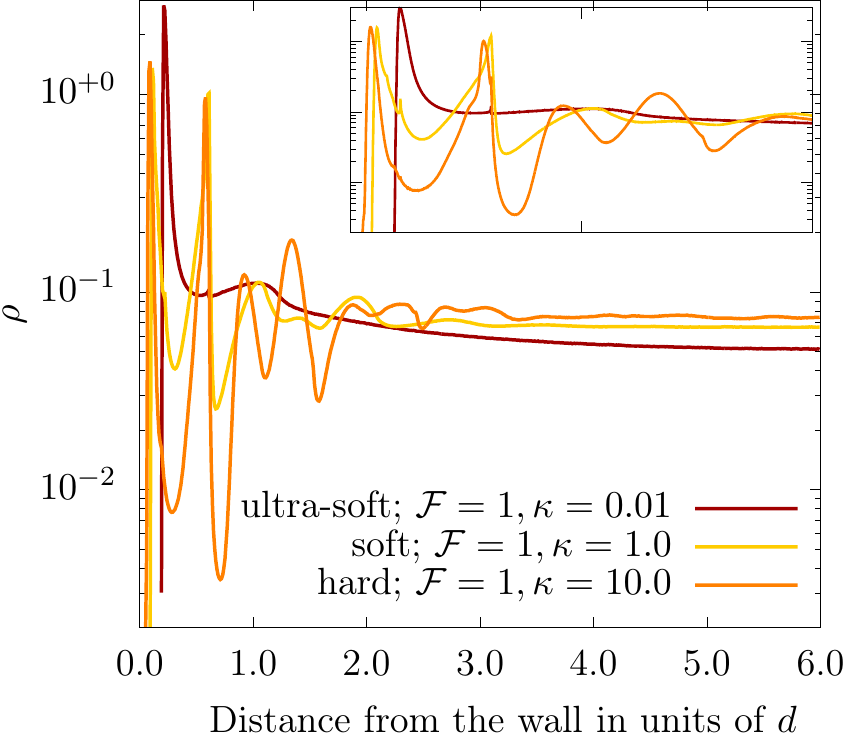}
\caption{
Rod density averaged over 100 realizations for rods in a channel. The inset contains a magnification of the region $\lambda \in [0:2]$.
}
\label{fig:homogeneous_density}
\end{center}
\end{figure}

For each realization, after the initial transient period we identify the center-of-mass ${\bf C}_{\alpha}$ for each rod and compute the rod density distribution (using the standard procedure \cite{allen2017computer, Heras2017}), given by: 

\begin{equation}
\rho(\bf{r}) = \langle \sum_{\alpha=1}^{N} \delta(\bf{r} - \bf{C}_{\alpha})\rangle_t,
\label{eq:density}
\end{equation}

To obtain an average density distribution we also average over a series of snapshots from a given run and over an ensemble of 100 realizations 
(where each realization is obtained by starting the system from a random initial configuration).
This gives us an averaged distribution as plotted in Fig \ref{fig:homogeneous_density}, in terms of distance from the wall $\lambda= \frac{W}{2}-|y|$.
The function is computed by dividing the channel into strips parallel to the channel axis and counting the number of rod centers within the strip. 
In addition, we normalize by $1/N$ to have comparable results in systems with different populations. 

\begin{figure}[h]
\begin{minipage}[h]{0.785\columnwidth}
\includegraphics[width=1.0\columnwidth]{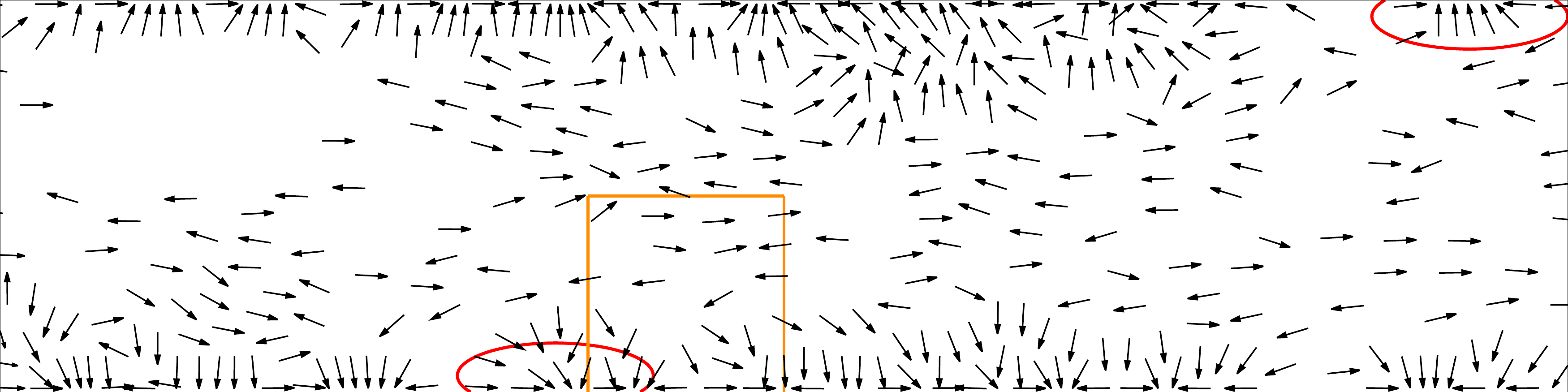}
\end{minipage}
\hfill
\begin{minipage}[h]{0.195\columnwidth}
\includegraphics[width=1.0\columnwidth]{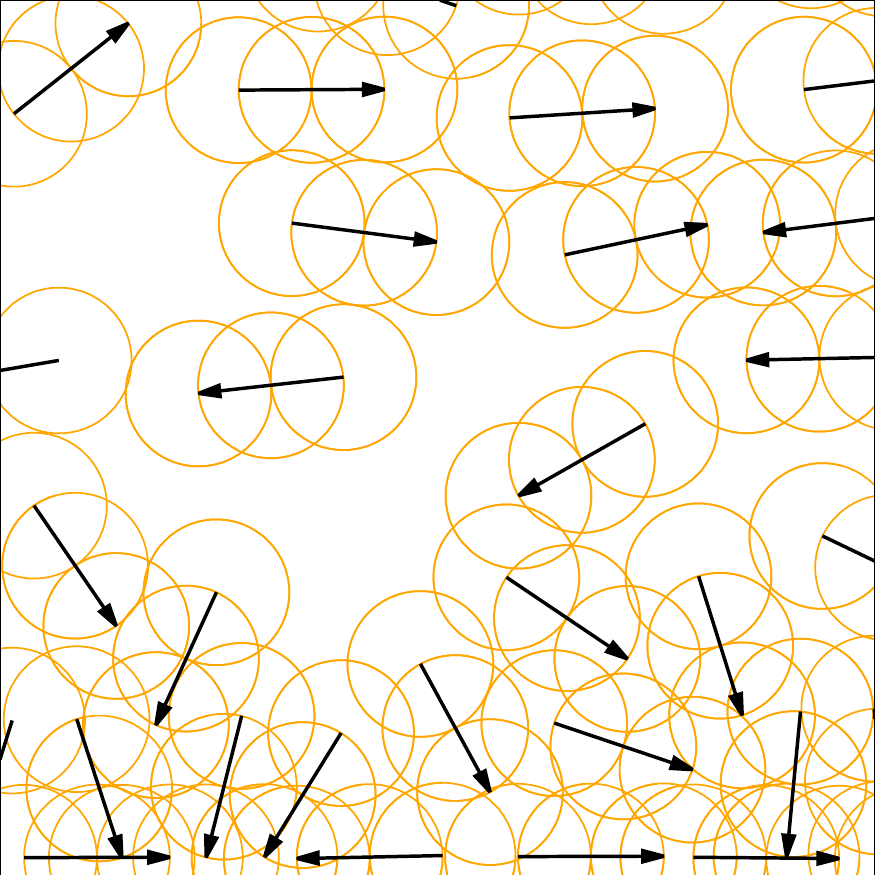}
\end{minipage}
\vfill
\begin{minipage}[h]{0.785\columnwidth}
\includegraphics[width=1.0\columnwidth]{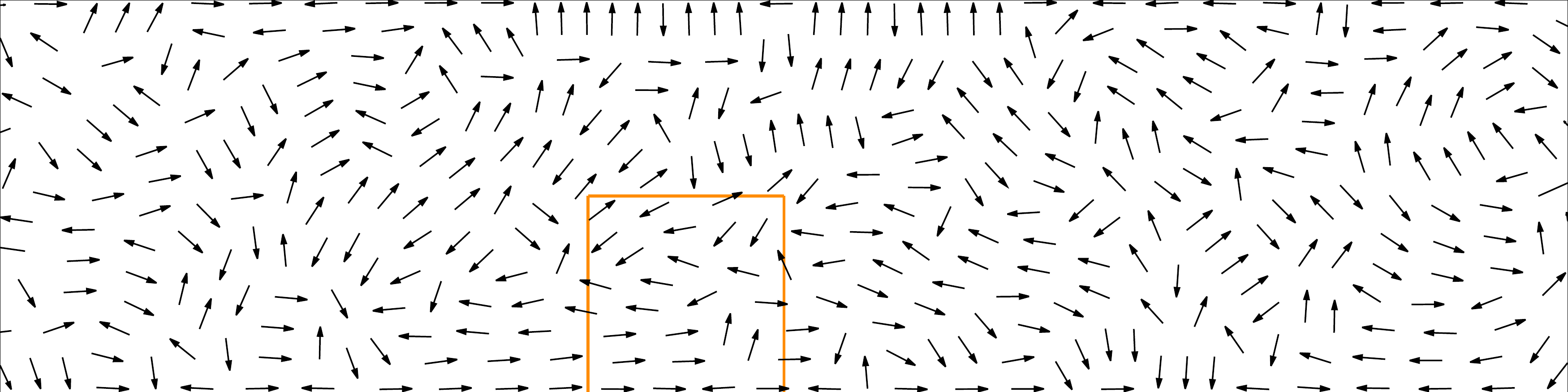}
\end{minipage}
\hfill
\begin{minipage}[h]{0.195\columnwidth}
\includegraphics[width=1.0\columnwidth]{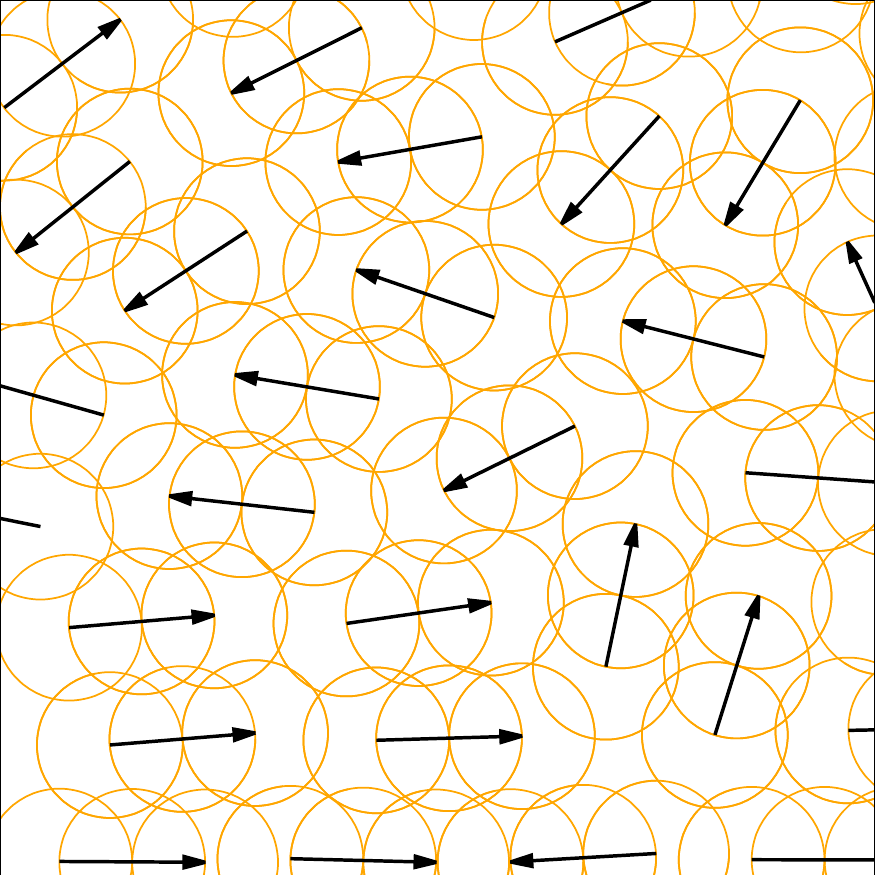}
\end{minipage}
\caption{
	Top: $\gamma=1.0$ (soft regime) and bottom $\gamma=10.0$ (hard regime).  \textit{Hedgehog}  structures are highlighted with (red) ellipses.
Right column contains magnified boundary regions with plotted segments. In the case of strong repulsive forces strong layering is observed.
(See Movies 1 and 2 in the Supplemental Material for representative animations of these two systems \cite{het_suppl}.)
}
\label{fig:snapshots}

\end{figure}

We find that for soft steric interactions, i.e., $\gamma=1$, most of the rods are concentrated at the boundaries of the channel, with a few rods in the interior (see yellow line in Fig. \ref{fig:homogeneous_density}, and the top image in Fig \ref{fig:snapshots}). 
This is apparent as a large peak in the rod density close to the boundary wall followed by a second layer.
After these surface layers the density is observed to drop rapidly.
In the limiting case of ultrasoft interactions (such as $\gamma=0.01$, as shown by the red line in Fig. \ref{fig:homogeneous_density}),
due to the extremely low repulsive forces almost all the rods are concentrated near the walls, tending to form a single boundary layer.

When steric interactions dominate (i.e., $\gamma = 10.0$) the rods are distributed more uniformly throughout the channel (see Fig. \ref{fig:homogeneous_density}, orange line, and bottom image in Fig \ref{fig:snapshots}).
In contrast to the soft systems, we observe a series of oscillations in the density going from the wall toward the interior indicating a layer like ordering.

The differences between the soft ($\kappa=1.0$) and hard ($\kappa=10.0$) regimes can be readily observed in a typical snapshot of the system. 
As shown in Fig \ref{fig:snapshots}, in soft systems the rods pile up at the boundaries forming short lived structures called \textit{hedgehogs} \cite{Marx2011,Wensink2008, Wioland2016},
while in the hard regime the rods are spread more evenly throughout the system and \textit{hedgehogs} are not observed.

\section{Heterogeneous populations in a periodic rectangular channel}\label{sec:hetero}
Our approach for heterogenous systems  is similar to that taken for the homogeneous cases described above in terms of simulation protocol, analysis, and numerical method. 

While in the case of homogeneous populations each rod in the system has the same hardness $\kappa$ and the same driving force $\mathcal{F}$, 
-- in the heterogeneous cases considered below we randomly assign one (or both) of these parameters for the population.
The cases we consider are: 
(i) keeping the hardness $\kappa$ the same for every rod while choosing a random value $\mathcal{F}$ for the self-propellant force (picked from a uniform distribution for each rod);
(ii) keeping the self-propellant force the same for each rod while assigning a random value $\kappa$ for the hardness.
Following this, in Sec. \ref{sec:full_hetero} we explore the problem of assigning a random value for $\kappa$ \emph{and} $\mathcal{F}$ to each rod in the population. 
We call these latter mixtures \emph{doubly heterogeneous} populations.

Our main finding is that in heterogeneous populations the presence of the channel boundaries leads to the demixing or segregation of rods. 
As in the homogeneous case where a strong layering effect is observed at the boundaries, so too in the heterogeneous populations we observe similar effects. 
However, here the rods are stratified into a sequence of layers, where the properties of the rods within a layer are roughly the same.

While the problem of mixed populations has been studied by some authors, they have mostly examined the case of binary mixtures. 
Here, two species with different motilities have been found to demix in semiperiodic channels \cite{Costanzo2014a}:
The fast moving rods are found in greater concentrations near the confining walls, while the slow rods are expelled to the interior. 
We show that this trend holds true even for populations where the motilities and hardness of the constituent rods are chosen from a continuous range of values.

\subsection{Heterogeneous populations with randomly assigned self-propellant forces}\label{sec:spf}

The first case is where the magnitude of the self-propellant force, for each rod, is assigned a random value picked from a uniform distribution $\mathcal{F}_{\alpha}\in [1,2)\quad\forall\alpha$.
The hardness of all the rods in the population is assigned a single value $\kappa$, with sets of experiments generated with different values of $\kappa$. 

\begin{figure}[!h]
\centering
\includegraphics[width=1.0\columnwidth]{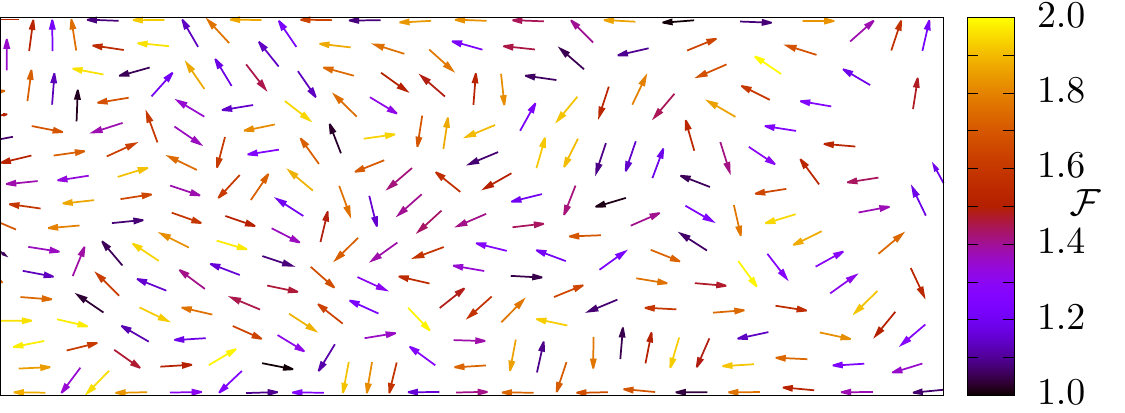}
\caption{
A snapshot of a system with heterogeneous self-propellent forces. 
All active rods are represented by arrows colored according to the values of their assigned self-propellant forces; 
the color bar  shows the range of $\mathcal{F}$ magnitudes, whilst $\kappa=10.0$ in this case.
(See Movie 3 in the Supplemental Material for a representative animation of this system \cite{het_suppl}.)
}
\label{fig:het_spf_example}
\end{figure}

Figure \ref{fig:het_spf_example} shows a representative snapshot of a system (with $\kappa=10.0$ and randomly assigned self-propellant forces) after the transient phase of the simulation. 
Fast and slow moving rods are colored yellow and purple, respectively, a color bar shows the range of $\mathcal{F}_{\alpha} $ values between these two extremes.

We refer to Secs. \ref{sec:theory} and \ref{sec:hom_systems} for general details of the simulation procedure. 
Here we note that every time we simulated the system each rod started with its own random initial configuration (position and orientation) 
\emph{and} a randomly assigned value for the self propellant force. 
In the first case we set $\kappa=1.0$ (i.e., soft interactions) and ran 100 simulations. 
This provided an ensemble on which we base our averaged results (as described below). 
Following this we ran the same experiment with $\kappa=10.0$ (i.e., hard interactions).

\begin{figure}[!h]
\begin{center}
\centering
\includegraphics[width=1.0\columnwidth]{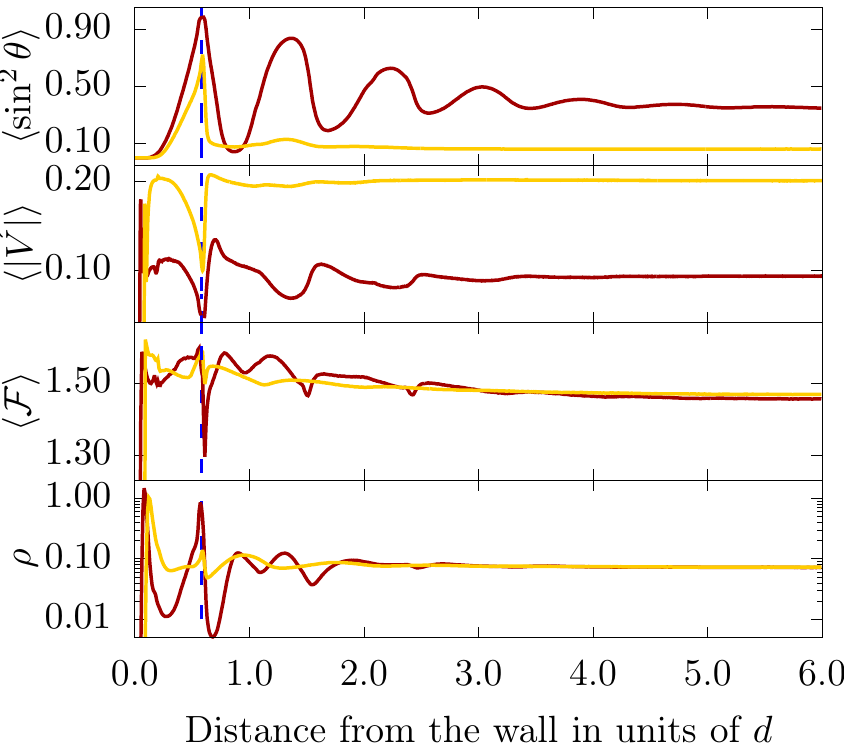}
\caption{
From top to bottom: average order parameter $\sin^2 \theta$, average speed $|\vec{V}|$, average self-propellant force $\mathcal{F}$, average density $\rho$;
$\sin^2\theta=0$ means parallel to walls, whereas $\sin^2\theta=1$ means perpendicular.
Red and yellow curves denote systems with $\kappa=10.0$ ($\gamma\approx 6.67$) and $\kappa=1.0$ ($\gamma\approx 0.67$), respectively.
The self propellant force for each rod is picked from a uniform distribution as described in the text. 
The system consists of a dense layer of rods at the channel walls followed by subsequent layers in the interior (the second one is highlighted by a vertical dashed line).
}
\label{fig:spf_plots_combined}
\end{center}
\end{figure}

The question we are interested in answering is the following: 
Do rods segregate so that the strongly driven rods are found in greater concentrations toward the boundaries? 
To answer this we split the channel into a series of narrow strips (along the length of the channel) and for some quantity of interest $\mathcal{A}$
(e.g., the self-propellant force of the rods $\mathcal{F}$, their orientation $\theta$, density $\rho$, and speed $|\vec{V}|$) we compute the average value of that quantity in the strip:
\begin{equation}
\langle\mathcal{A}\rangle(\bf{r})=\frac{\langle \sum_{\alpha=1}^{N}\mathcal{A}_{\alpha} \delta(\bf{r}-\bf{C}_{\alpha}) \rangle_t}{\langle \sum_{\alpha=1}^{N}\delta(\bf{r}-\bf{C}_{\alpha}) \rangle_t}.
\label{eq:spf_distribution}
\end{equation}
An analysis of the systems with $\kappa=1.0$ and $\kappa=10.0$ is summarised in Fig. \ref{fig:spf_plots_combined}. 


We interpret Fig. \ref{fig:spf_plots_combined} as follows. 
The plot of the density distribution $\rho$ shows a large concentration of rods at the boundaries where $\lambda\approx 0$ regardless of the value of $\kappa$.
These surface rods are orientated \textit{parallel} to the walls (as inferred from the the orientation parameter $\langle\sin^2 \theta \rangle$) 
and are on average more strongly driven than the rods in the interior (as shown by plotting $\langle \mathcal{F} \rangle$) --- similar to previous studies of mixed populations \cite{Fily2014a}.
Thus, a universal feature of these systems is that strongly driven rods are found at the channel boundaries 
as the result of an \emph{expelling} process \cite{Fily2014a} whereby weakly driven rods are pushed out of the surface layer.
However, the distribution of rods is also sensitive to the choice of $\kappa$, so that the layerlike ordering is less pronounced in softer systems. 
We also observe a slight decrease in the average motility of the rods, i.e., $\langle \mathcal{F} \rangle$, at the boundaries as we increase the steric forces between rods.

Adjacent to the surface layer we observe a second spike in the density at $\approx0.5\lambda$ (blue dashed line in Fig. \ref{fig:spf_plots_combined}),
which is particularly sharp when steric interactions are strong. 
This is generated by rods which are also in contact with the wall but are orientated \emph{perpendicular} to it (as inferred from the plot of the orientation parameter). 
This second layer consists of rods which are weakly driven (compared to first layer) and are trapped against the channel wall.
Since they are perpendicular to the wall they do not contribute to actively pushing the surface layer of of rods. 
These rods are unable to reorientate themselves due to strong \emph{caging} \cite{Yang2014a} and are dragged along by the faster rods in the surface layer. 
Subsequent layers, particularly for hard systems, show some of the features of the two outer layers,
but it becomes increasing difficult to distinguish distinct layers as we go into the interior of the system. 

\subsection{Heterogeneous populations with randomly assigned repulsive coefficients}\label{sec:kappa}
\begin{figure}[h]
\begin{center}
\centering
\begin{minipage}[h]{1\linewidth}
\includegraphics[width=1.0\columnwidth]{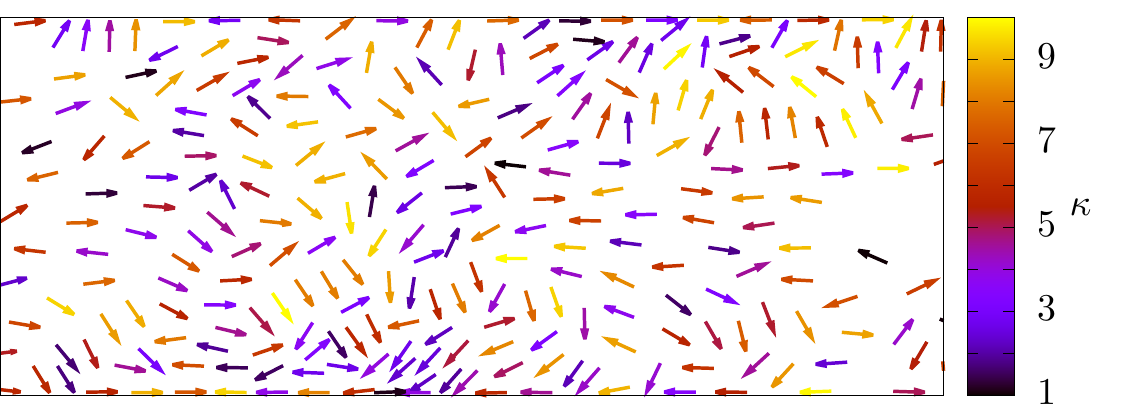}
\end{minipage}
\caption{
A snapshot of a system with heterogeneous softness. 
All active rods are represented by arrows colored according to the values of their assigned hardness; 
the color bar shows the range of $\kappa$ magnitudes, while $\mathcal{F}=1.0$ in this case. 
(See Movie 4 in the Supplemental Material for a representative animation of this system \cite{het_suppl}.)
}
\label{fig:kappa_snap}
\end{center}
\end{figure}
We now consider the inverse situation whereby every rod has its own randomly assigned value of $\kappa$ (taken from the uniform distribution, $\kappa\in[1.0,10.0)$, 
but the magnitude of the self propellant force $\mathcal{F}$ for each rod in the population is the same, with sets of experiments generated with different values of $\mathcal{F}$.

\begin{figure}
\begin{center}
\centering
\includegraphics[width=1.0\columnwidth]{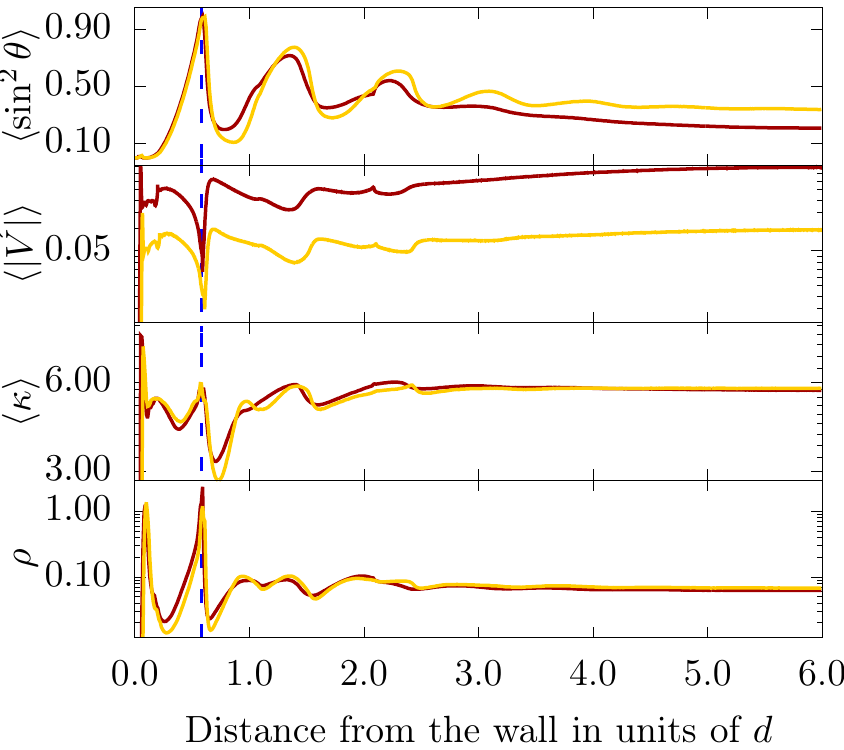}
\caption{
From top to bottom: average order parameter $\sin^2 \theta$, average speed $|\vec{V}|$, average hardness $\kappa$, average density $\rho$. 
Red curves denote systems with $\mathcal{F}=2.0$ ($\gamma=2.75$), yellow curves with $\mathcal{F}=1.0$ ($\gamma=5.5$). 
The hardness of the rods is picked from a uniform distribution as described in the text.
The first of internal layers is indicated by the vertical dashed line.}
\label{fig:kappa_plots_combined}
\end{center}
\end{figure}

To compute the interaction between rods with different hardnesses we introduce a Lorentz-Berthelot  \cite{Lorentz1881} combining rule for an effective potential.
For any two rods $\alpha$ and $\beta$ the average $\kappa$ for the interaction between them, is given by,
\begin{equation}
\kappa_{\alpha\beta}=\sqrt{\kappa_{\alpha} \kappa_{\beta}}.
\end{equation}
Figure \ref{fig:kappa_snap} shows a representative snapshot for a system with $\kappa \in[1.0,10.0)$ and $\mathcal{F}=1.0$. 
Soft and hard rods are colored blue and yellow, respectively, while a color bar indicates rods with hardnesses intermediate between these two extremes. 

As before we plot the averaged quantities of the system in Fig. \ref{fig:kappa_plots_combined} and find that the system is composed of a series of layers. 
The outer layer with $\lambda\approx 0.0$ is composed on average of the hardest rods. 
Surprisingly, the average velocity of the rods in this layer is relatively high compared to the rest of the system. 
Next to the outermost layer is an internal layer of slower moving softer rods which are perpendicular to the walls. 
Immediately behind these surface layers is a region containing very soft rods which are locally trapped into dense clusters.

\section{Doubly heterogeneous systems in a rectangular channel}\label{sec:full_hetero}

The final case is where both the hardness and the self-propellant force for each rod is assigned from a random distribution. 
We use the same distributions as in the previous cases, that is $\kappa \in[1.0,10.0)$ and $\mathcal{F} \in  [1,2)$, resulting in $\gamma\approx 3.67$.

\begin{figure}[!h]
\begin{minipage}[h]{1\linewidth}
\includegraphics[width=1.0\columnwidth]{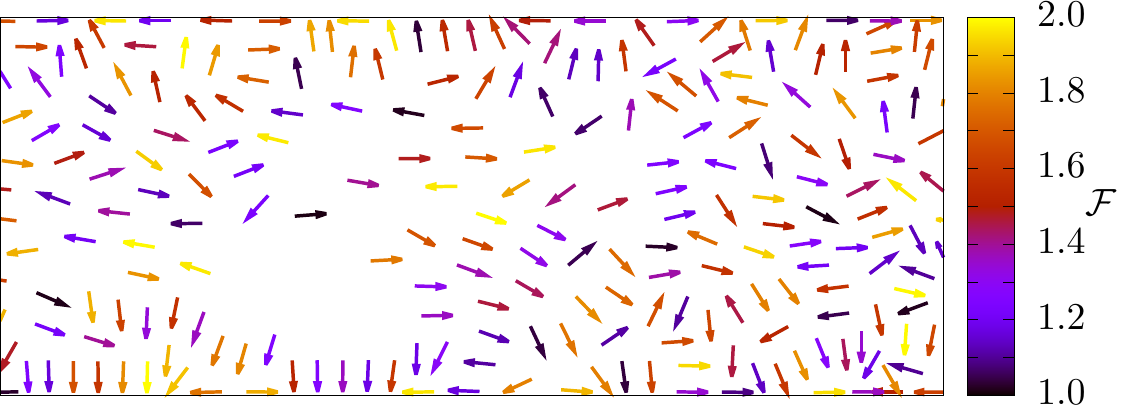}
\end{minipage}
\hfill
\begin{minipage}[h]{1\linewidth}
\includegraphics[width=1.0\columnwidth]{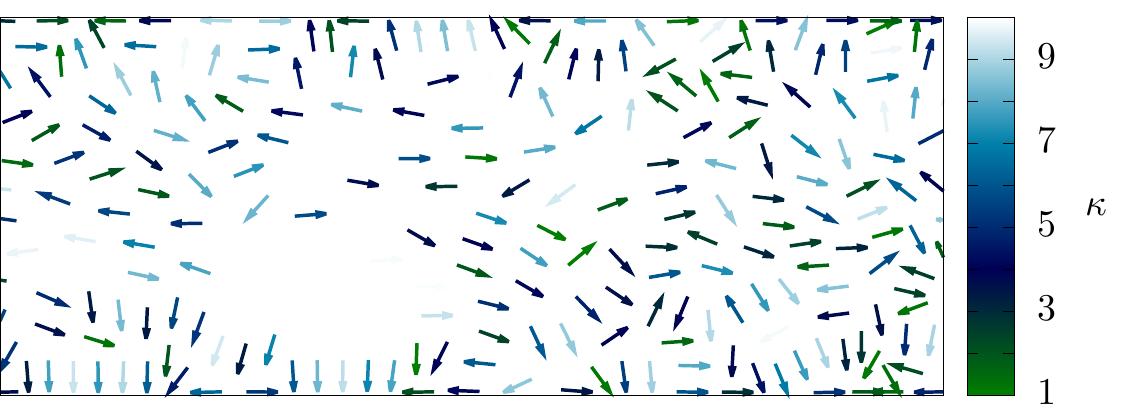}
\end{minipage}
\caption{
Snapshots of a doubly heterogeneous system. 
Top: rods are colored according to their values of $\mathcal{F}$; bottom: rods are colored according to their values of $\kappa$.
(See Movies 5 and 6 in the Supplemental Material for representative animations of this system \cite{het_suppl}.)
}
\label{fig:full_snapshots}
\end{figure}

A representative snapshot of the system is shown in Fig. \ref{fig:full_snapshots}, 
where the top figure shows the rods colored according to their assigned self propellant force and the bottom figure shows the same rods colored according to hardness.

\begin{figure}[!h]
\begin{center}
\centering
\includegraphics[width=1.0\columnwidth]{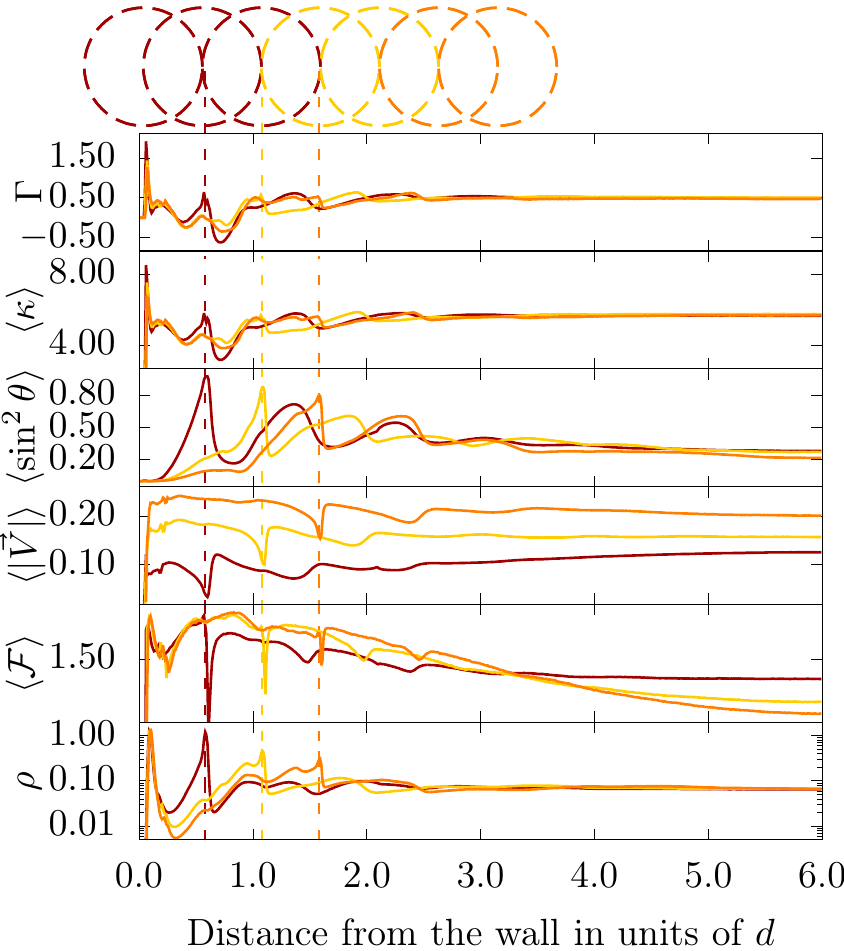}
\caption{
Distributions in doubly mixed systems, from top to bottom: $\Gamma$ defined by Eq. (\ref{eq:Gamma}), 
average hardness $\kappa$, average order parameter $\sin^2 \theta$, average speed $|\vec{V}|$, average self-propellant force $\mathcal{F}$, and average density $\rho$. 
Values of $\kappa$ and $\mathcal{F}$ of the rods are picked from uniform distributions as described in the text. 
Red curves denote systems where rods are composed of three segments, yellow and orange stand for systems with five and seven segments, respectively.
The first of the internal layers is indicated by the vertical dashed line of corresponding color. Subsequent layers are also highlighted by dashed lines.
The disks at the top of the figure represent rods of different lengths aligned perpendicularly to the wall, corresponding to the peaks in the graphs.
}
\label{fig:mixed_distributions}
\end{center}
\end{figure}

The behavior of the system can again be understood with reference to the various quantities plotted in Fig. \ref{fig:mixed_distributions}. 
This doubly mixed system shows features from both the previous cases. 
The system is again composed of dense layers at the boundaries, 
but this time we find that the outermost layer consists of rods which are on average highly motile in terms of $\mathcal{F}$ \emph{and} the hardest. 
In the particular case studied here, we observe evidence of subsequent layers but these decay quickly to give way to a disordered interior. 

To ensure that these results are not specific to a single aspect ratio (that is length of rod versus its width) 
we show the outcome for simulations for a range of rod lengths (specifically rods composed of three, five, and seven segments in length). 
In all cases we find there is an outer layer, composed of rods parallel to the walls, 
followed by a second layer at a distance half the length of the rod (i.e., composed of rods at the wall boundaries which are perpendicular to the walls).

To show that active rods at the boundaries are \textit{both} motile and hard, we introduce a metric:
\begin{equation}
\Gamma(\bf{r}) = \frac{\langle \sum_{\alpha=1}^{N}(\mathcal{F}_{\alpha}-\bar{\mathcal{F}})(\kappa_{\alpha}-\bar{\kappa}) \delta(\bf{r}-\bf{r}_{\alpha})\rangle_t}
{\bar{\mathcal{F}}\bar{\kappa} \langle \sum_{\alpha=1}^{N}\delta(\bf{r}-\bf{r}_{\alpha})\rangle_t},
\label{eq:Gamma}
\end{equation}
where $\bar{\mathcal{F}}$ and $\bar{\kappa}$ are the average motility and hardness of the rods, respectively. 
Thus, positive $\Gamma$ corresponds to rods that are either hard with high self-propellant forces or soft with low self-propellant forces. 
Figure \ref{fig:mixed_distributions} shows that $\Gamma$ has a strong peak right next to the wall. 
Furthermore, because both $\langle \kappa\rangle$ and  $\langle \mathcal{F} \rangle$ are high in this region
we deduce that the layer of surface rods is (on average) formed by hard rods with high self-propellant forces.
\section{Conclusions}
\label{sec:conclusions}
We have studied the properties of a swarm of actively driven rods in a rectangular channel. 
We have compared the case of a homogeneous swarm to that of a heterogeneous swarm (whereby one or more of the dynamical features are randomly assigned). 

In the case of the heterogeneous systems our key finding is that the channel wall drives the segregation of the population. 
Thus rods which are hard and fast moving are more likely to be found at the edge of the system, while soft and less motile rods are pushed in to the interior. 
We have deliberately confined these simulations to narrow channels and it remains to be seen if these results apply to wider systems in which vortex type behavior is commonly seen.

While these features have been hinted at by previous simulations of binary mixtures of rods, 
here we demonstrate that segregation is present even in systems where the dynamical properties are given by a continuous range --- as is the case for many bacterial communities. 

An important test case for these findings maybe the work of Ilkanaiv \textit{et al.} \cite{Ilkanaiv2017}, where swarming \textit{Bacillus subtilis} move on surfaces. 
In this experimental study the rodlike bacteria have their lengths and motilities  distributed heterogeneously. 
For such populations we expect to see segregation if the bacteria are in the presence of a wall or confined within a channel.

While the system studied here is idealised model of some active systems (e.g., bacterial populations),
nevertheless we hope that these findings might hint at a passive means of sorting active populations according to their dynamical properties. 
Indeed, these findings already suggest that boundaries can be used to enhance the concentration of rods with high motilities and strong steric interactions.
The design of channel geometries that can then syphoned off these rods from the rest of the population (and their efficiency as sorting devices) will be the subject of future publications.

\begin{acknowledgments}
	Both V.K. and A.M. acknowledge support from MPNS COST (European Cooperation in Science and Technology) Action MP1305 ``Flowing matter''. Both V.K. and A.M. acknowledge numerous useful discussions with Hartmut L{\"o}wen and his group. We acknowledge the use of Supercomputing Wales. 
\end{acknowledgments}
\bibliography{library}

\end{document}